# AN ULTRA-FAST METHOD FOR GAIN AND NOISE PREDICTION OF RAMAN AMPLIFIERS

*Ann M Rosa Brusin[1]\*, Vittorio Curri[1], Darko Zibar[2], Andrea Carena[1]*

[1]*Dipartimento di Elettronica e Telecomunicazioni, Politecnico di Torino, Torino, Italy*
[2]*DTU Fotonik, Technical University of Denmark, Lyngby, Denmark*
*\*ann.rosabrusin@polito.it*

## Abstract

A machine learning method for prediction of Raman gain and noise spectra is presented: it guarantees high-accuracy (RMSE < 0.4 dB) and low computational complexity making it suitable for real-time implementation in future optical networks controllers.

## 1 Introduction

Optical amplification schemes exploiting the Stimulated Raman Scattering (SRS) are currently experiencing a revival because of their ability to provide gain and low-noise figure at any wavelength, up to the entire O+E+S+C+L bands, which are considered for the next generation of optical communication systems [1]. Fast routing, deployment and optimization of data traffic will be highly demanded, as network automatization at low-latency is highly desired in the path toward autonomous and self-adaptive optical networks. Therefore, ultra-fast methods for predicting gain and noise profiles for Raman amplification are essential. The standard approach is to solve a system of nonlinear ordinary differential equations (ODE) governing forward and backward propagation of optical signals spectra in presence of SRS. However, this approach is time-consuming and computationally demanding, especially when considering a large number of pumps needed to enable amplification in wide-band systems.

The use of machine learning (ML) in optical communications has widespread in recent years targeting different applications [2,3]. Concerning the analysis of Raman amplifiers (RA), the main focus has been on the application of machine learning techniques for pump allocation to obtain the desired gain [4,5]. A study based on ML techniques to predict RA gain and noise profiles was published in [6], with a single specific method and a modest validation set. In our present work, we target the same goal proposing an approach based on multi-layer neural networks (NN), comparing alternative training algorithms and activation functions. Moreover, we optimize NN hyper-parameters and we carry out a comprehensive validation over a very large number of conditions. After training on a data set, a neural network can give excellent predictions for gain and noise profiles but it is several orders of magnitude faster than the standard approach based on the ODE solver as it only relies on matrix multiplications. Therefore, it is suitable for real-time implementation. In our study we compare the two most popular training algorithms for learning the weights in neural networks, back-propagation [7] and random projection methods [8]. The latter, beside requiring lower computational time for the training phase, is able to predict gain and noise profiles with very limited RMSE, always lower than 0.4 dB.

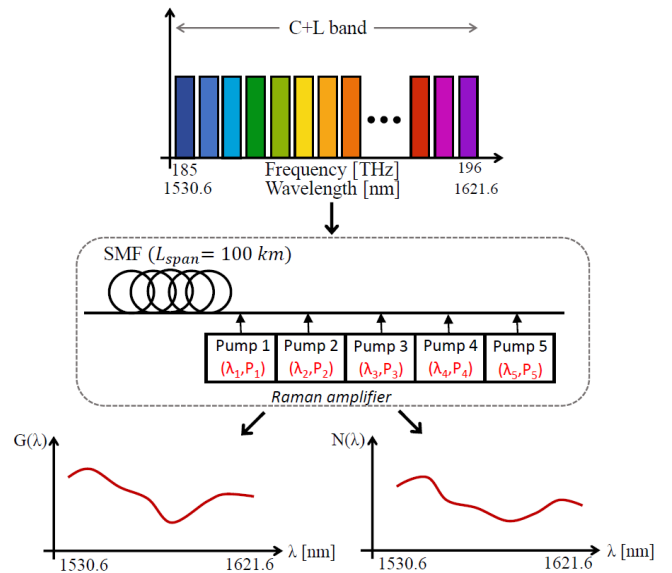

Fig. 1 Schematic of a single span Raman amplifier using five counter-propagating pumps.

## 2. Simulation set-up and machine learning framework

We consider a single span RA (Fig 1), with five counter-propagating pumps ($[\lambda_i, P_i]$ with $i = 1, …,5$), and evaluate Raman gain $G(\lambda)$ and noise $N(\lambda)$ profiles. In our study, we consider the C+L band (11 THz from 185 THz to 196 THz, i.e. from 1530.6 nm to 1621.6 nm) with a resolution bandwidth $B_W = 100$ GHz. The input to the RA is a Wavelength Division Multiplexing (WDM) comb of 343 channels, each operated with polarization-division multiplexed coherent technologies at the symbol rate of 32 GBaud, Nyquist shaped and with 0 dBm power, loaded in the whole C+L band, see Fig 1. A single span ($L_{span} = 100$ km) Single-Mode Fibre (SMF) is considered with the following fibre parameters: attenuation $\alpha_S$ = 0.2 dB/km for the signals and $\alpha_P$ = 0.25 dB/km for the pumps, chromatic dispersion $D$ = 16.7 ps/nm/km, effective area $A_{eff}$ = 80 µm$^2$, non-linear coefficient $\gamma$ = 1.26 1/W/km and Raman coefficient $c_R$ = 0.4125 1/W/km.

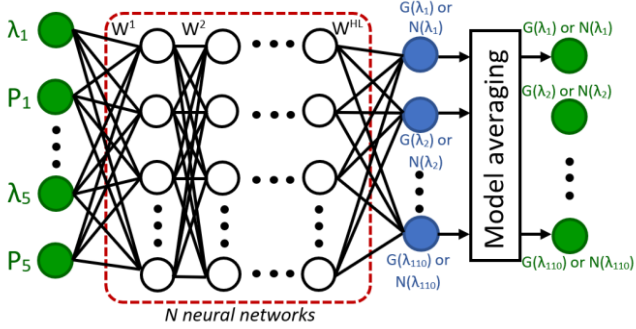

Fig. 2 General representation of a multi-layer neural network. We perform model averaging over $N$ parallel and independent neural networks to determine the mapping between the pump wavelengths and powers and the corresponding Raman gains $G(\lambda)$ or noise figures $N(\lambda)$.

We use two independent NNs, one for learning the mapping from the pumps wavelengths and powers [$\lambda_1$, $P_1$, $\lambda_2$, $P_2$, $\lambda_3$, $P_3$, $\lambda_4$, $P_4$, $\lambda_5$, $P_5$] to the gain $G = [G(\lambda_1),...,G(\lambda_{110})]$ and the second one for learning the mapping to the noise $N = [N(\lambda_1),...,N(\lambda_{110})]$. The general structure of the employed NNs is shown in Fig 2. To improve the performance of NNs in terms of predictions, we run $N$ independent and parallel neural networks and we average their output.

For the training of the NNs, two different algorithms, back-propagation (BP) and random projection (RP), are implemented to learn the weight matrices [$\mathbf{W}^1,...,\mathbf{W}^{HL}$], which connect the input layer to the hidden layers (HL), and then to the output layer (Fig 2).

Using the ODE solver, we generate a data set, with $M$ = 5000 elements, drawing pump wavelengths and power from uniform distributions: $\lambda_1^i \sim U[1424, 1436.2]$ nm, $\lambda_2^i \sim U[1436.2, 1458.4]$ nm, $\lambda_3^i \sim U[1458.4, 1480.6]$ nm, $\lambda_4^i \sim U[1480.6, 1502.8]$ nm and $\lambda_5^i \sim U[1502.8, 1525]$ nm, $P_1^i, P_2^i, P_3^i, P_4^i, P_5^i \sim U[0, 160]$ mW to guarantee a complete coverage of the range of gains of practical interest.

When using BP, the learning algorithm is the Levenberg-Marquardt, the number of hidden layers is 2, the number of hidden nodes is 10 and the number of parallel and independent NNs over which we perform model averaging is $N$ = 10. We also performed model selection by investigating different activation functions such as hyperbolic tangent and the logistic sigmoid.

When the RP method is used, we consider a Single-hidden Layer Feed-forward Neural Network (SLFN), such that number of hidden layers is 1, and model averaging is computed over $N$ = 20 parallel and independent NNs. Also, in this case we perform model selection trying different activation functions: sine, hyperbolic tangent and logistic sigmoid. For each of them and for both gain and noise profile predictions, we search for the optimal number of hidden nodes, sweeping it from 20 to 600 with step of 20. We found that the optimal values of number of hidden nodes are 400, 120 and 240

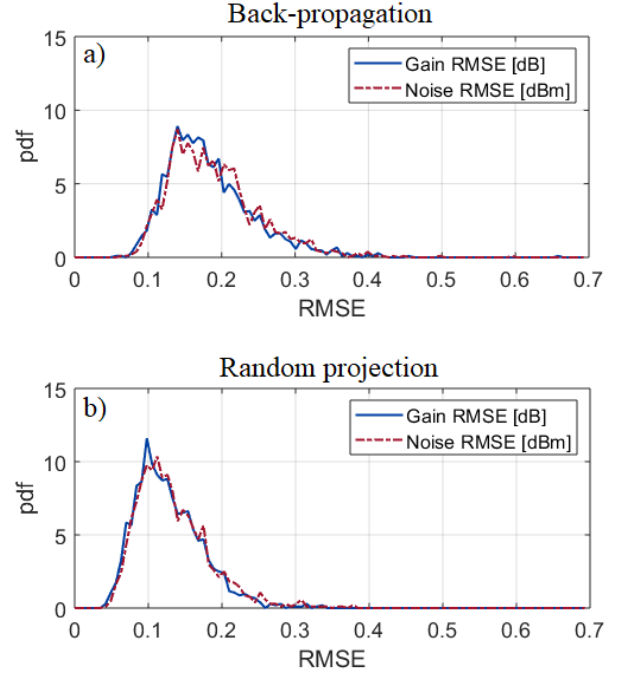

Fig. 3 Probability density function (pdf) of the RMSE for the predicted Raman gain and noise profiles in case of: (a) BP and hyperbolic tangent activation function and (b) RP and sine activation function.

respectively for sine, hyperbolic tangent and logistic sigmoid activation functions.

Even if it does impact the computational time of the NN when in operation, we must report a significative difference between BP and RP in the training phase: the required time is up to three order of magnitude in favour of RP. In fact, whether for RP the training of a single NN is instantaneous, for BP it requires a time of the order of hours.

## 3   Numerical results and accuracy analysis

To investigate if the trained NNs, can accurately predict the gain and noise profiles on the unseen data we use a second independent data set (test stage). As we want to validate the trained NNs in condition that could be practical, we prune the data set, selecting only cases where minimum and maximum values of gain in the profile are inside the range from 4 to 12 dB. We assume that below 4 dB is not worth to implement a RA for such a low gain, while 12 dB is the threshold to remain in the Moderated Pumping Regime (60% of 20 dB span loss) [9] where a RA is more convenient and it also avoids saturation effects.

The performance of the NN is evaluated by defining the prediction errors as:

$$\Delta \mathbf{G}(\lambda) = \mathbf{G}_{pred}(\lambda) - \mathbf{G}_{target}(\lambda) \qquad (1)$$

$$\Delta \mathbf{N}(\lambda) = \mathbf{N}_{pred}(\lambda) - \mathbf{N}_{target}(\lambda) \qquad (2)$$

where $\mathbf{G}_{pred}(\lambda)$ and $\mathbf{N}_{pred}(\lambda)$ are the NN predicted profiles and $\mathbf{G}_{target}(\lambda)$ and $\mathbf{N}_{target}(\lambda)$ are the target profiles evaluated using the ODE solver.

For each element of the validation data set, we evaluate the root-mean-square-error (RMSE) and maximum absolute error ($Error_{MAX}$), over the whole C+L band, of the prediction errors

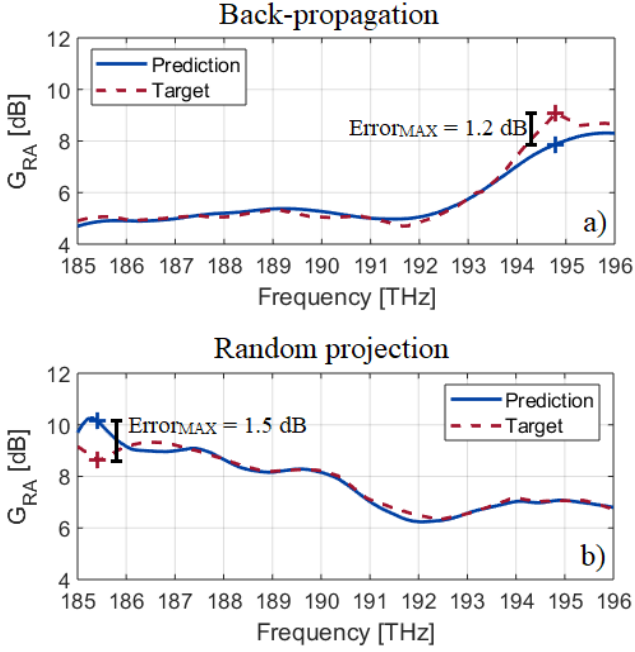

Fig. 4 Predicted and corresponding target gain profiles in the worst case using (a) BP and (b) RP. Worst case means that we select the profiles for which the prediction error is maximum among the other predicted profiles.

$\Delta G(\lambda)$ and $\Delta N(\lambda)$ defined in (1) and (2). Among the investigated activation functions, we report results only for those giving the best RMSE for each NN solution. Fig 3 shows the probability density functions (pdfs) of the RMSE both of gains and of noise profiles. In Fig 3a we show results for the case of BP and hyperbolic tangent activation function (tanh). Whereas Fig 3b illustrates the same quantities but when RP and sine activation function are considered. Comparing the two training methods, the shape of the pdf is similar but slightly steeper for RP than BP. Moreover, gain and noise RMSEs resulting from BP, have both higher mean value (respectively 0.19 dB and 0.19 dBm) and standard deviation (respectively 0.06 dB and 0.06 dBm) than those obtained in case of RP (0.13 dB and 0.14 dBm means, 0.05 dB and 0.05 dBm standard deviations), meaning that the predictions are also slightly more accurate when RP is used. A further proof that RP is better than BP is in the fact that it provides a maximum value of RMSE of 0.34 dB, for gains, and of 0.38 dBm, for noise, lower than values obtained in case of BP, which are respectively 0.66 dB and 0.60 dBm. RMSE is a good parameter to measure the quality of the prediction over the whole C+L bandwidth, but to guarantee that the proposed NNs are not affected by local errors over narrow band regions, we considered also the maximum absolute prediction error. In the worst case, we observed a 1.2 dB maximum error between prediction and target profile in case of BP (Fig 4a) and 1.5 dB in case of RP (Fig 4b). We are aware that such values are not negligible, but as it can be seen in Fig 4, the prediction error impacts only a small frequency region. To further understand the likelihood of incurring in a large gain prediction error, we analysed its distribution over the validation set.

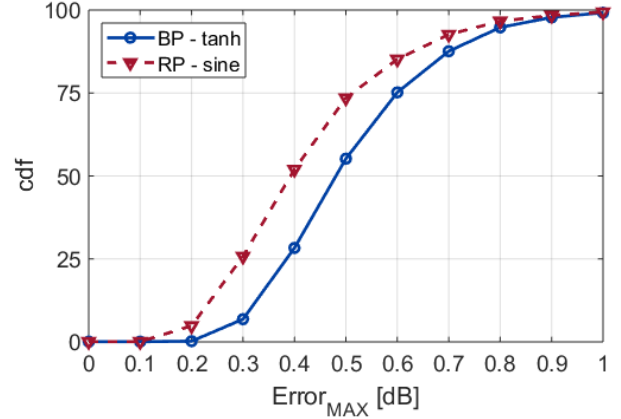

Fig. 5 Cumulative distribution function (cdf) of the maximum absolute gain error $Error_{MAX}$ for both BP and RP trained NNs.

Fig 5 shows the cumulative distribution function (cdf) of the maximum error where we can read for example, that using BP more than 50% of the cases shows a maximum error below 0.5 dB and more than 99.4% are below 1 dB. Results are even better in case of RP, since about 75% of cases have a maximum error below 0.5 dB, whilst the percentage of errors above 1 dB is similar to BP. Similar results are obtained when we predict noise profiles.

## 4 Conclusion

It has been numerically demonstrated that machine learning offers significant advantages for predicting Raman amplification gain and noise profiles in terms of speed and computational complexity. A maximum prediction error below 0.6 dB over the whole C+L band for more than 75% of cases has been demonstrated making it an attractive solution for integrated network controllers for next generation optical networks.
From our study, the RP approach shows an advantage over BP because of the reduced computational time needed for training. We tested the proposed method in a highly demanding condition, C+L bands with 5 pumps, but same principles can further be scaled up to an even higher number of pumps to cover also other bands. The analysis we have shown here for SMF fibre and a span length of 100 km can be extended to other fibre types and span lengths, expecting the same level of prediction accuracy.
Moreover, to avoid modelling approximations and parameter identification uncertainties, the whole approach presented in this paper, where training has been based on an artificial data set, can be applied using an experimental data set to train the NN. Under these conditions, we expect an improvement in the accuracy of predictions for the practical operation of the Raman amplifier.

## 5 Acknowledgements

This work is supported by the European Research Council through the ERC-CoG FRECOM project (grant agreement no. 771878).